\documentclass[fleqn,preprint,showpacs,preprintnumbers,amsmath,amssymb]{revtex4}
\usepackage{graphicx}% Include figure files
\usepackage{dcolumn}% Align table columns on decimal point
\usepackage{bm}% bold math
\begin{document}

\title{Global limits on kinetic Alfv\'{e}non speed in quasineutral plasmas}
\author{M. Akbari-Moghanjoughi}
\affiliation{Azarbaijan University of
Tarbiat Moallem, Faculty of Sciences,
Department of Physics, 51745-406, Tabriz, Iran}

\date{\today}
\begin{abstract}
Large amplitude kinetic Alfv\'{e}non (exact Alfv\'{e}n soliton) matching condition is investigated in quasineutral electron-ion and electron-positron-ion plasmas immersed in a uniform magnetic field. Using the standard pseudopotential method, the magnetohydrodynamics (MHD) equations are exactly solved and a global allowed matching condition for propagation of kinetic solitary waves is derived. It is remarked that, depending on the plasma parameters, the kinetic solitons can be sub- or super-Alfv\'{e}nic, in general. It is further revealed that, either upper or lower soliton speed-limit is independent of fractional plasma parameters. Furthermore, the soliton propagation angle with respect to that of the uniform magnetic field is found to play a fundamental role in controlling the soliton matching speed-range.
\end{abstract}


\maketitle

\section{Background}

In the collective processes in plasma if the magnetic field perturbations are also taken into account, in addition to the ion-acoustic mode, the Alfv\'{e}n mode will also be present. Hence, the plasma will be dispersive if the kinetic effects like ion-drift velocity are also included in the ion continuity and the ion momentum equations. In fact, it is because of inclusion of these effects that the propagating waves are named the kinetic waves. In other words, when the perpendicular wavelength is comparable to the ion gyro-radius, the ions will no longer follow the magnetic field lines of force, while the electrons/positrons, due to their small Larmor-radius, will be attached to the magnetic force lines. This causes an effective charge separation which as a result will lead to the kinetic Alfv\'{e}n waves. Therefore, the ion drift-velocity is mainly responsible for the Alfv\'{e}n waves kinetic effects. On the other hand, kinetic Alfv\'{e}n solitary waves occur due to the interplay between the dispersive behavior and nonlinear steepening effects \cite{roy}.

Nonlinear Alfv\'{e}n waves such as Alfv\'{e}nons (exact Alfv\'{e}n soliton)and shocks play important role in solar and astrophysical plasmas \cite{louran, dovn}. Stasiewicz \cite{stas} has recently suggested that Alfv\'{e}nons can provide an explanation for various aspects of electromagnetic energy dissipation and heating in the solar corona and in planetary magnetospheres. Early investigations of Alfv\'{e}nons date back to 1976 \cite{hasegawa, Yu}. Application of exact methods such as the Sagdeev approach is a rigorous tool in the study of the nonlinear phenomena \cite{vedenov, davidson, sagdeev} in plasmas, which provides vital information concerning the wave dynamics. There has been many recent studies, mostly using the Sagdeev method, on the possibility of solitary Alfv\'{e}n waves in diverse plasmas with dust and positron contaminant, considering the plasma inhomogeneities and various charge distribution effects both in low-beta and inertial limits \cite{das, wu, kaka, wang, anup, chen, rajku, ansar, saleem, mirza, lyang, cheong, yang, tao, rub, masood, run}.

Investigations show that, both rarefactive and compressive solitary waves exist depending on the fractional plasma parameters and the value of plasma-beta ($\beta=2\mu_0 n_{i0} k_B T_e/B_0^{2}$), a parameter which in most cases is assumed to be smaller than unity. Furthermore the theoretical studies of kinetic Alfv\'{e}nons reveals the existence of both sub- and super-Alfv\'{e}nic nonlinear structures. In a recent study O. P. Sah \cite{sah} has reported the existence of super-Alfv\'{e}nic double-layers in electron-positron-ion plasmas. Woo etal. \cite{woo} have found that the collisional effects can be important in converting the soliton-shape structures to double-layers in dusty plasmas. Prasanta Chatterjee etal. \cite{pras}, have recently studied the effect of ion-temperature on large-amplitude kinetic solitary waves and concluded that the ion temperature has important effect on shape of solitary kinetic Alfv\'{e}n waves and double-layers. Current investigation is an attempt to study the most general cases of electron-ion and electron-positron-ion three dimensional plasmas consisting of warm electrons/positron and ions considering the two dimensional drifts of ions. We will report features in arbitrary amplitude soliton dynamics which is common in all electron-ion and electron-positron-ion plasmas regardless of the charge distribution function for plasma species. The presentation of the article is as follows. In Sec. \ref{bas} the basic fluid formalism is presented. The general pseudo-potential solution and the required conditions on the solitary propagations is given in Sec. \ref{sul}. We extend our methodology to the electron-positron cases in Sec. \ref{epi} and give several examples in Sec. \ref{ex}

\section{Magnetohydrodynamics Model}\label{bas}

We use the conventional magnetohydrodynamics (MHD) formalism for a three-dimensional homogenous and collisionless quasineutral plasma present in a uniform magnetic field, $\bm{B}=B_0\bm{\hat{e_z}}$, which is directed along the $z$-axis. In this formalism two potentials, namely, $\varphi$ and $\psi$, describe parallel and perpendicular perturbations in the electric field \cite{kadomtsev}, $\bm{E}$, respectively,. Furthermore, we consider only kinetic perturbations in the magnetic field. We also assume that the ions possess both polarization drift and $\bm{E}\times \bm{B}$ velocity components in $x$-$y$ plane. Hence, the set of normalized MHD fluid equations read as \cite{anup}
\begin{equation}\label{dimensional}
\begin{array}{l}
{\partial _t}{n_e} + {\partial _\parallel}\left( {{n_e}{u_{\parallel e}}} \right) = 0, \\ {\partial _t}{n_i} + {\nabla _ \bot }\cdot\left( {{n_i}{{\bf{u}}_{\perp i}}} \right) + {\partial _\parallel}\left( {{n_i}{u_{\parallel i}}} \right) = 0, \\ {{\bf{u}}_{\perp i}} = {{\bf{E}}_ \bot } \times {{{\bf{\mathord{\buildrel{\lower3pt\hbox{$\scriptscriptstyle\frown$}}
\over e} }}}_{\bf{\parallel}}} + {d_t}{{\bf{E}}_ \bot },\hspace{3mm}{{\bf{E}}_ \bot } =  - {\nabla _ \bot }\psi ,\hspace{3mm}{\nabla _ \bot } \equiv \left( {{\partial _x},{\partial _y}} \right), \\ Q\left[ {{\partial _t}{u_{\parallel e}} + {u_{\parallel e}}{\partial _\parallel}{u_{\parallel e}}} \right] - {\partial _\parallel}\varphi  + n_e^{ - 1}{\partial _\parallel}{P_e}({n_e}) = 0, \\ {\partial _t}{u_{\parallel i}} + \left( {{{\bf{u}}_{\perp i}} \cdot {\nabla _ \bot }} \right){u_{\parallel i}} + {u_{\parallel i}}{\partial _\parallel}{u_{\parallel i}} + {\partial _\parallel}\varphi  + n_i^{ - 1}{\partial _\parallel}{P_i}({n_i}) = 0, \\ {\partial _\parallel}\left[ {{\Delta _ \bot }\left( {\psi  - \varphi } \right)} \right] = \left( {\beta /2} \right){\partial _t}{J_\parallel},\hspace{3mm}{\Delta _ \bot } \equiv \left( {\partial _x^2,\partial _y^2} \right),\hspace{3mm}{J_\parallel} = {n_i}{u_{\parallel i}} - {n_e}{u_{\parallel e}}, \\
\end{array}
\end{equation}
where, $n_{j}$, $u_{j}$, $m_{j}$, $P_j$ and $J$ are density, velocity, mass, pressure of $j$-th species ($j=e,i$) and the total current density, respectively. Furthermore, the notations $\parallel$ and $\perp$ refer to the parallel and perpendicular (with respect to direction of the magnetic field) vector components. The parameter $\beta=2\mu_0 n_{i0} k_B T_e/B_0^{2}$ ($n_{i0}$ being the equilibrium ion number density) is the ratio of thermal to magnetic pressure and the quantity $Q$ is the electron to ion mass ratio to be neglected in our analysis. In obtaining the normalized equation set (Eqs. (\ref{dimensional})) we have used the general scalings
\begin{equation}\label{nm}
\begin{array}{l}
\{x,y,z\} \to \frac{c_s}{\omega_{ci}} \{\bar x,\bar y,\bar z\},\hspace{2mm}t \to \frac{{\bar t}}{{{\omega _{ci}}}},\hspace{2mm}{n_{j}} \to {n_{i0}}{\bar n_{j}}, \hspace{2mm}{\bf{u_j}} \to {c_s}{\bf{\bar u_j}},\\ \varphi  \to \frac{\epsilon}{e}\bar \varphi,\hspace{2mm}\psi  \to \frac{\epsilon}{e}\bar \psi,\hspace{2mm} P_{j} \to  {\epsilon} \bar P_{j},\hspace{2mm}J \to  en_{i0}c_s \bar J,
\end{array}
\end{equation}
where, $\omega_{ci}=eB_0/m_i$, $c_s=\sqrt{\epsilon/m_i}$ are ion cyclotron-frequency and acoustic speed, respectively. The value of parameter $\epsilon$ will be defined based on the electronic charge distribution.

\section{Kinetic Solitons in Electron-ion Plasmas}\label{sul}

Being interested in stationary wave solutions moving at constant velocity, we reduce Eqs. (\ref{dimensional}) to co-moving stationary soliton frame by changing into the new coordinate $\xi=l_x x + l_y y + l_z z - M t$ ($l_x^{2}+l_y^{2}+l_z^{2}=1$), with $M=V/V_A$ being the normalized matching soliton-speed, where $V_A=B_0/\sqrt{\mu_0 n_{i0} m_i}$ is the Alfv\'{e}n wave-speed. By changing the coordinate, making use of the quasineutrality condition, $n_e=n_i$, and integrating with appropriate boundary conditions ($\mathop {\lim }\limits_{\xi  \to \infty} \varphi = 0,\hspace{3mm}\mathop {\lim }\limits_{\xi  \to \infty} {\bf{u_j}} = {\bf{0}},\hspace{3mm}\mathop {\lim }\limits_{\xi  \to \infty} {n_j} = 1$), the reduced set of equations become
\begin{equation}\label{red}
\begin{array}{l}
{l_x}{u_{xi}} + {l_y}{u_{yi}} + {l_z}{u_{zi}} = M(1 - {n^{ - 1}})\\
{u_{ze}} = l_z^{ - 1}M(1 - {n^{ - 1}}) \\
{u_{xi}} = M{l_x}{\partial _{\xi \xi }}\psi  - {l_y}{\partial _\xi }\psi  \\
{u_{yi}} = {l_x}{\partial _\xi }\psi  + M{l_y}{\partial _{\xi \xi }}\psi  \\
{u_{zi}} = {l_z}{M^{ - 1}}n\Phi^{*} (n),\hspace{3mm}\Phi^{*} (n) = \int_1^n {{n^{ - 1}}{d_{n}}P(n)dn} , \\
{l_z}\left( {l_x^2 + l_y^2} \right)\left( {{\partial _{\xi \xi }}\psi  - {\partial _{\xi \xi }}\varphi } \right) = n\beta M\left( {{u_{ze}} - {u_{zi}}} \right)/2,
\end{array}
\end{equation}
where, $P(n)=P_e(n)+P_i(n)$ and $\Phi^{*} (n)$ is the effective plasma potential due to the total pressure. The following general differential equation is obtained by combining Eqs. (\ref{red})
\begin{equation}\label{sol}
\frac{{{\partial ^2}\varphi (n)}}{{\partial {\xi ^2}}} =\frac{1}{{1 - l_z^2}}\left[ {\frac{{\beta {M^2}}}{{2l_z^2}}(1-n) + \frac{\beta{n^2}}{2}{\Phi ^*}(n) - \frac{{l_z^2n}}{{{M^2}}}{\Phi ^*}(n)} + (1 - {n^{ - 1}}) \right].
\end{equation}
Algebraic manipulation and integration with the aforementioned boundary conditions, results in the well-known energy integral of the form
\begin{equation}\label{energy}
\frac{1}{2}{\left( {\frac{{dn}}{{d\xi }}} \right)^2} + U(n) = 0, \\
\end{equation}
with the desired pseudo-potential
\begin{equation}\label{pseudo}
U(n) = \frac{{{{\left[ {{d_n}\varphi (n)} \right]}^{ - 2}}}}{{1 - l_z^2}}\int_1^n {\frac{{d\varphi (n)}}{{dn}}\left[ {\frac{{\beta {M^2}}}{{2l_z^2}}(n - 1) - \frac{{\beta {n^2}}}{2}{\Phi ^*}(n) + \frac{{l_z^2n}}{{{M^2}}}{\Phi ^*}(n) + ({n^{ - 1}} - 1)} \right]dn}.
\end{equation}
where
\begin{equation}\label{w}
\varphi(n) = \int_1^n {{n^{ - 1}}{d_{n}}P_e(n)dn}.
\end{equation}
The possibility of solitary excitation relies on some conditions to satisfy, simultaneously, namely
\begin{equation}\label{conditions}
{\left. {U(n)} \right|_{n = 1}} = {\left. {\frac{{dU(n)}}{{dn}}} \right|_{n = 1}} = 0,\hspace{3mm}{\left. {\frac{{{d^2}U(n)}}{{d{n^2}}}} \right|_{n = 1}} < 0.
\end{equation}
It is further required that for at least one either maximum or minimum nonzero $n$-value, we have $U(n_{m})=0$, so that for every value of $n$ in the range ${n _m} > n  > 1$ (compressive soliton) or ${n _m} < n  < 1$ (rarefactive soliton), $U(n)$ is negative (it is understood that there is no root in the range $[1,n_m]$). In such a condition we can obtain a potential minimum which describes the possibility of a solitary wave propagation. The stationary soliton solutions corresponding to this pseudo-potential which satisfies the mentioned boundary-conditions, read as
\begin{equation}\label{soliton}
\xi  - {\xi _0} =  \pm \int_1^{n_m} {\frac{{dn}}{{\sqrt { - 2U(n)} }}}.
\end{equation}
The conditions for the existence of a solitary propagation stated above require that, first, it takes infinitely long pseudo-time ($\xi$) for the system to get away from the unstable point ($n=1$). This statement requires that $d_n U(n)\mid_{n=1}=0$ or equivalently $d_\xi n\mid_{\xi=-\infty}=0$ in parametric space, as it is also inferred by the shape of a solitary wave. Thereafter, moving forward in pseudo-time ($\xi$) axis, the localized density perturbation reaches a maximum or a minimum at $n=n_m$ (if it exists) at which the pseudo-speed ($d_\xi n$) of the analogous particle bound in pseudo-potential ($U(n)$) region of $1>n>n_m$ (or $1<n<n_m$) reaches zero again and it returns back. Note that, in the parametric space, from equation Eq. (\ref{soliton}), it is observed that in physical situation $U(n)$ should be negative for solitary (non-periodic) wave solution, which is clearly satisfied if $d_{nn}U(n)\mid_{n=1}<0$ and $U(n_m\neq 1)=0$. Note also that, both the requirements $U(n)\mid_{n=1}=0$ and $d_{n}U(n)\mid_{n=1}=0$ follow from the equilibrium state assumption at infinite pseudo-time ($\xi=\pm\infty$) before and after perturbation takes place, i.e. $d_{\xi\xi} n\mid_{\xi=\pm\infty}=d_{\xi} n\mid_{\xi=\pm\infty}=0$. However, there is a special case with $d_{n}U(n)\mid_{n=n_m}=0$ for which the density perturbation is stabilized at the maximum or minimum density $n=n_m$ (the analogous particle never returns back). This situation regards to the existence of a double-layer in plasma which is not considered here. \textbf{Extra requirement such as $(n_m-1)d_{\xi\xi}n\mid_{\xi=0}<0$ given in Ref. \cite{wu2} is needed for exclusion of the possibility of double layer or shock-like structure in addition to the ones presented in Eq. (\ref{conditions}). However, in the present analysis we only attempt to give a general criterion for the soliton matching speed with the assumption that such density profile exists for the set of given plasma parameters in Eq. (\ref{pseudo}). Therefore, we continue to evaluate the matching condition with the assumptions that a root $n_m\neq 1$ exists and $(n_m-1)d_{\xi\xi}n\mid_{\xi=0}<0$ (to exclude shock-like solutions).} Moreover, the pseudopotential given by Eq. (\ref{pseudo}) and its first derivative vanish at $n=1$, as required by the two first conditions in Eq. (\ref{conditions}). Also, direct evaluation of the second derivative of the Sagdeev potential, Eq. (\ref{pseudo}), at unstable point, $n=1$, leads to
\begin{equation}\label{con2}
{\left. {\frac{{{d^2}U(n)}}{{d{n^2}}}} \right|_{n = 1}} = \frac{{l_z^2{{\left[ {{d_n}\varphi (1)} \right]}^{ - 1}}}}{{1 - l_z^2}}\left[ {\frac{M}{{{l_z}}} - \sqrt {{d_n}{\Phi ^*}(1)} } \right]\left[ {\frac{M}{{{l_z}}} - \sqrt {\frac{2}{\beta }} } \right],\hspace{3mm}{d_n}{\Phi ^*}(1) = {\left. {{d_n}P(n)} \right|_{n = 1}}.
\end{equation}
Thus, in order for the existence of oblique kinetic Alfv\'{e}n solitary propagations, the matching soliton-speed must satisfy the following general inequalities
\begin{equation}\label{con1}
\left\{ {\begin{array}{*{20}{c}}
{\sqrt {{d_n}{\Phi ^*}(1)}  < \frac{M}{{{l_z}}} < \sqrt {\frac{2}{\beta }} ;} & {\beta {d_n}{\Phi ^*}(1) < 2}  \\
{\sqrt {\frac{2}{\beta }}  < \frac{M}{{{l_z}}} < \sqrt {{d_n}{\Phi ^*}(1)} ;} & {\beta {d_n}{\Phi ^*}(1) > 2}  \\
\end{array}} \right\}
\end{equation}
This condition is analogous to the one presented for oblique electrostatic solitary propagations in magnetoplasmas given in Ref. \cite{akbari1}. However, in current case the propagation angle with respect to the magnetic field, $l_z$, plays a fundamental role in controlling the matching speed range. Note that there is also a critical $\beta$-value ($\beta_{cr} = 2 \left[{d_n}{\Phi ^*}(1)\right]^{-1}$) at which no kinetic Alfv\'{e}nons can propagate. Furthermore, no kinetic Alfv\'{e}nons are allowed to propagate at right angle to the magnetic field.

\section{Extension to Electron-Positron-Ion Plasmas}\label{epi}

We extend the methodology used above to obtain matching conditions for electron-positron-ion kinetic Alfv\'{e}n solitary propagations. To this end, we change to the new coordinate $\xi=l_x x + l_y y + l_z z - M t$ ($l_x^{2}+l_y^{2}+l_z^{2}=1$), use appropriate boundary conditions $\mathop {\lim }\limits_{\xi  \to \infty } \varphi = 0$ and $\mathop {\lim }\limits_{\xi  \to \infty } n_{i,e,p} = \{1,\alpha,\alpha-1\}$, where $\alpha=n_{e0}/n_{i0}\geq 1$ (the special case of $\alpha=1$ leads to the quasineutral electron-ion magnetoplasma), and employ the quasineutrality condition, $n_i=n_e-n_p$, to obtain similar relations as Eqs. (\ref{red})
\begin{equation}\label{normalp}
\begin{array}{l}
{l_x}{u_{xi}} + {l_y}{u_{yi}} + {l_z}{u_{zi}} = M(1 - {n_i^{ - 1}})\\
{n_e}{u_{ze}} = l_z^{ - 1}M({n_e}-\alpha) \\
{n_p}{u_{zp}} = l_z^{ - 1}M({n_p}-\alpha + 1) \\
{u_{xi}} = M{l_x}{\partial _{\xi \xi }}\psi  - {l_y}{\partial _\xi }\psi  \\
{u_{yi}} = {l_x}{\partial _\xi }\psi  + M{l_y}{\partial _{\xi \xi }}\psi  \\
{u_{zi}} = {l_z}{M^{ - 1}}n_i\Psi^{*} (n_i),\hspace{3mm}\Psi^{*} (n_i) = \varphi(n_i) + \int_1^{n_{i}} {{n_i^{ - 1}}{d_{n_i}}P_i(n_i)dn_i} , \\
{l_z}\left( {l_x^2 + l_y^2} \right)\left( {{\partial _{\xi \xi }}\psi  - {\partial _{\xi \xi }}\varphi } \right) = \beta M\left( {n{}_e{u_{ze}} - {n_p}{u_{zp}}} - {n_i}{u_{zi}} \right)/2,
\end{array}
\end{equation}
where, the function $\varphi(n_i)$ is given by the following relations
\begin{equation}\label{con}
\frac{{\partial \varphi }}{{\partial n_e }} = \frac{1}{{{n_e}}}\frac{{\partial {P_e}({n_e})}}{{\partial n_e }},\hspace{3mm}\frac{{\partial \varphi }}{{\partial n_p }} = - \frac{1}{{{n_p}}}\frac{{\partial {P_p}({n_p})}}{{\partial n_p }},\hspace{3mm}{n_e} - {n_p} \approx {n_i},{\rm{ }}.
\end{equation}
Using the above relations, we expect to find the function $\varphi=\varphi(n_i)$ which obviously is nontrivial in most cases. However, as it will be revealed, we need to calculate the quantity $d_{\varphi}n_i(\varphi)$, instead, which is feasible in many cases. Therefore, for the current case, Eq. (\ref{sol}) changes to
\begin{equation}\label{sol2}
\frac{{{\partial ^2}\varphi (n_i)}}{{\partial {\xi ^2}}} =\frac{1}{{1 - l_z^2}}\left[ {\frac{{\beta {M^2}}}{{2l_z^2}}(1-n_i) + \frac{\beta{n_i^{2}}}{2}{\Psi ^*}(n_i) - \frac{{l_z^2n_i}}{{{M^2}}}{\Psi ^*}(n_i)} + ({n_i}-1) \right].
\end{equation}
The generalized Sagdeev pseudopotential $U(n_i)$, then, reads as
\begin{equation}\label{pseudo2}
\frac{{{{\left[ {{d_{{n_i}}}\varphi ({n_i})} \right]}^{ - 2}}}}{{1 - l_z^2}}\int_1^{{n_i}} {\frac{{d\varphi ({n_i})}}{{d{n_i}}}\left[ {\frac{{\beta {M^2}}}{{2l_z^2}}({n_i} - 1) - \frac{{\beta n_i^2}}{2}{\Psi ^*}({n_i}) + \frac{{l_z^2{n_i}}}{{{M^2}}}{\Psi ^*}({n_i}) + (1 - {n_i})} \right]d{n_i}}.
\end{equation}
The matching conditions to be evaluated is of the similar shape as before
\begin{equation}\label{con2}
\left\{ {\begin{array}{*{20}{c}}
{\sqrt {{d_{n_i}}{\Psi ^*}(1)}  < \frac{M}{{{l_z}}} < \sqrt {\frac{2}{\beta }} ;} & {\beta {d_{n_i}}{\Psi ^*}(1) < 2}  \\
{\sqrt {\frac{2}{\beta }}  < \frac{M}{{{l_z}}} < \sqrt {{d_{n_i}}{\Psi ^*}(1)} ;} & {\beta {d_{n_i}}{\Psi ^*}(1) > 2}  \\
\end{array}} \right\},
\end{equation}
however, with the new definition of the effective potential given below
\begin{equation}
{d_{{n_i}}}{\Psi ^*}(1) = {\left. {{{\left. {{{\left[ {{d_\varphi }{n_i}(\varphi )} \right]}^{ - 1}}} \right|}_{\varphi  = 0}} + {d_{{n_i}}}{P_i}({n_i})} \right|_{{n_i} = 1}}.
\end{equation}

\section{Case Study for Soliton Matching Condition}\label{ex}

It clearly remarked from Eqs, \ref{con1} and \ref{con2} that both sub- and super-Alfv\'{e}non may exist depending on the equation of state for electrons/positrons and ions. Considering the pressure of ions to be $P_i(n_i)=n_i^{\gamma} k_B T_i/\epsilon$, where $\gamma=(f+2)/f$ is the adiabatic constant and $f$ is the ion degrees of freedom (DoF) and with the cases of $\gamma=1$ for isothermal-ions and $\gamma\neq 1$ for adiabatic-ions, we employ different electronic distributions to evaluate the possibility condition for solitary propagations. For instance for the Maxwell-Boltzmann, Lorentzian (Kappa), Thomas-Fermi and Fermi-Dirac \cite{akbari2} electron/positron distributions (appropriately normalized), we have
\begin{equation}\label{n}
\left[ {\begin{array}{*{20}{c}}
{{\rm{Maxwell}}} & {\left\{ {\begin{array}{*{20}{c}}
{{n_e} = \alpha {e^\varphi }}  \\
{{n_p} = (\alpha  - 1){e^{ - \delta \varphi }}}  \\
\end{array}} \right\}} & { \epsilon= {k_B}{T_e}} & {\delta  = \frac{{{T_p}}}{{{T_e}}}}  \\
{{\rm{Lorentz}}} & {\left\{ {\begin{array}{*{20}{c}}
{{n_e} = \alpha {{(1 - \varphi /(k - 3/2))}^{ - k + 1/2}}}  \\
{{n_p} = (\alpha  - 1){{(1 + \delta \varphi /(k - 3/2))}^{ - k + 1/2}}}  \\
\end{array}} \right\}} & { \epsilon= {k_B}{T_e}} & {\delta  = \frac{{{T_p}}}{{{T_e}}}}  \\
{{\rm{Fermi}}} & {\left\{ {\begin{array}{*{20}{c}}
{{n_e} = \alpha {{(1 + \varphi )}^{3/2}}}  \\
{{n_p} = (\alpha  - 1){{(1 - \delta \varphi )}^{3/2}}}  \\
\end{array}} \right\}} & { \epsilon= {k_B}{T_{Fe}}} & {\delta  = \frac{{{T_{Fp}}}}{{{T_{Fe}}}}}  \\
{{\rm{Dirac}}} & {\left\{ \begin{array}{l}
\varphi  = \sqrt {1 + {\alpha ^{2/3}}{n_e}^{2/3}\eta _0^2}  - \sqrt {1 + {\alpha ^{4/3}}\eta _0^2}  \\
\varphi  = \sqrt {1 + {{(\alpha  - 1)}^{4/3}}\eta _0^2}  - \sqrt {1 + {{(\alpha  - 1)}^{2/3}}{n_p}^{2/3}\eta _0^2}  \\
\end{array} \right\}} & { \epsilon= {m_e}{c^2}} & {\eta _0^3 = \frac{{{n_{i0}}}}{{{N_0}}}}  \\
\end{array}} \right].
\end{equation}
where, $\alpha$ is the electron-to-ion equilibrium number-density ratio and $T_{Fe,p}$ is the electron/positron Fermi-temperature, respectively. The parameter $\eta_0$ is related to the plasma mass-density (of white dwarf, for instance, where the relativistic degeneracy occurs) through the relation $\rho\simeq 2m_p n_{0}$ or $\rho(gr/cm^{3})=\rho_0 \eta_0^{3}$ with $\rho_0(gr/cm^{3})\simeq 1.97\times 10^6$, where, $m_p$ is the proton mass and the cases $\bar\rho(=\rho/\rho_0)\ll 1$ and $\bar\rho\gg 1$ correspond to the nonrelativistic and ultrarelativistic degeneracy limits, respectively \cite{akbari2}. The density $\rho_0$ is exactly in the range of a mass-density of a typical white dwarf (the density of typical white dwarfs can be in the range $10^{5}<\rho(gr/cm^{3})<10^{9}$). The normalized matching Alfv\'{e}non-speed in the case of electron-positron-ion magnetoplasma is bounded through inequality of the form, Eq. (\ref{con2}), with the values of ${d_{n_i}}\Psi^* (1)$ given as
\begin{equation}\label{con8}
\left[ {\begin{array}{*{20}{c}}
{{\rm{Maxwell}}} & {{d_{{n_i}}}{\Psi ^*}(1) = \gamma \sigma  + {{\left[ {\alpha  + \delta (\alpha  - 1)} \right]}^{ - 1}}} & {} & {\sigma  = \frac{{{T_i}}}{{{T_e}}}}  \\
{{\rm{Lorentz}}} & {{d_{{n_i}}}{\Psi ^*}(1) = \gamma \sigma  + \frac{{2k - 3}}{{2k - 1}}{{\left[ {\alpha  + \delta (\alpha  - 1)} \right]}^{ - 1}}} & {} & {\sigma  = \frac{{{T_i}}}{{{T_e}}}}  \\
{{\rm{Fermi}}} & {{d_{{n_i}}}{\Psi ^*}(1) = \gamma \sigma  + \frac{2}{3}{{\left[ {\alpha  + \delta (\alpha  - 1)} \right]}^{ - 1}}} & {} & {\sigma  = \frac{{{T_i}}}{{{T_{Fe}}}}}  \\
{{\rm{Dirac}}} & {{d_{{n_i}}}{\Psi ^*}(1) = \frac{{{\eta _0} }}{{\sqrt {{{3(\alpha  - 1)}^{1/3}}\sqrt {1 + {{(\alpha  - 1)}^{4/3}}\eta _0^2}  + 3{\alpha ^{1/3}}\sqrt {1 + {\alpha ^{4/3}}\eta _0^2} } }}} & {} & {\sigma  \sim 0}  \\
\end{array}} \right],
\end{equation}
where, the $\alpha=1$ case corresponds to the electron-ion limit. For instance, the $\alpha=1$ (electron-ion) limiting case of the Kappa distribution in Eq. (\ref{con8}) has been investigated in Ref. \cite{run}. Note also that for the limit of spectral index $k\rightarrow\infty$ the result for the Maxwell-Boltzmann distribution is obtained. Evaluation of Eq. (\ref{con8}) indicates that the effects of adiabatic index, $\gamma$, and fractional ion-temperature, $\sigma$, is to shrink (widen) the solitary matching speed-range, while, the effects of fractional electron number-density, $\alpha$, and fractional positron temperature, $\delta$, is to widen (shrink) the corresponding range for $\beta {d_{n_i}}{\Psi ^*}(1) < 2$ ($\beta {d_{n_i}}{\Psi ^*}(1) > 2$). It is also observed from the values given in Eq. (\ref{con8}) that, the increase in the plasma mass-density in a Fermi-Dirac relativistically degenerate plasma shrinks (widens) the Mach-number range for the case $\beta {d_{n_i}}{\Psi ^*}(1) < 2$ ($\beta {d_{n_i}}{\Psi ^*}(1) > 2$).

\end{document}